\documentclass{PoS}

\title{New CTEQ Global Analysis with High Precision Data from the LHC}

\ShortTitle{New CTEQ Global Analysis: CT18(Z) PDFs}

\author{Tie-Jiun Hou\\
	Department of Physics, Northeastern University, Shenyang, Liaoning, China\\
	E-mail: \email{tjhou@msu.edu}}

\author{Keping Xie\\
        Department of Physics, Southern Methodist University, Dallas, TX 75275-0181, U.S.A.\\
        E-mail: \email{kepingx@mail.smu.edu}}
\author{Jun Gao\\
	INPAC, Shanghai Key Laboratory for Particle Physics and Cosmology \\ \& School of Physics and Astronomy, Shanghai Jiao Tong University, Shanghai 200240, China \\
	Center for High Energy Physics, Peking University, Beijing 100871, China. \\
	E-mail: \email{jung49@sjtu.edu.cn}}
\author{Sayipjamal Dulat\\
	School of Physics Science and Technology,
	Xinjiang University, Urumqi, Xinjiang 830046, China\\
	E-mail: \email{sdulat@hotmail.com}}
\author{Marco Guzzi\\
	Department of Physics, Kennesaw State University, 370 Paulding Ave.,  30144 Kennesaw, GA, U.S.A.\\
	E-mail: \email{mguzzi@kennesaw.edu}}
\author{T. J. Hobbs\\
Department of Physics, Southern Methodist University, Dallas, TX 75275-0181, U.S.A.\\
Jefferson Lab, EIC Center, Newport News, VA 23606, U.S.A.\\
	E-mail: \email{tjhobbs@mail.smu.edu}}
\author{Joey Huston\\
	Department of Physics and Astronomy, Michigan State University, East Lansing, MI 48824 U.S.A.\\
	E-mail: \email{Huston@pa.msu.edu}}
\author{Pavel Nadolsky\\
Department of Physics, Southern Methodist University, Dallas, TX 75275-0181, U.S.A.\\
	E-mail: \email{nadolsky@physics.smu.edu}}
\author{Jon Pumplin\\
	Department of Physics and Astronomy, Michigan State University, East Lansing, MI 48824 U.S.A.\\
	E-mail: \email{pumplin@pa.msu.edu}}
\author{Carl Schmidt\\
	Department of Physics and Astronomy, Michigan State University, East Lansing, MI 48824 U.S.A.\\
	E-mail: \email{schmidt@pa.msu.edu}}
\author{Ibrahim Sitiwaldi\\
	School of Physics Science and Technology,
Xinjiang University, Urumqi, Xinjiang 830046, China\\
	E-mail: \email{ibrahim010@sina.com}}
\author{Dan Stump\\
	Department of Physics and Astronomy, Michigan State University, East Lansing, MI 48824 U.S.A.\\
	E-mail: \email{stump@pa.msu.edu}}
\author{\speaker{C.-P. Yuan}\\
	Department of Physics and Astronomy, Michigan State University, East Lansing, MI 48824 U.S.A.\\
	E-mail: \email{yuan@pa.msu.edu}}

\abstract{We present the new CTEQ-TEA global analysis of quantum chromodynamics (QCD). In this analysis, parton distribution functions (PDFs) of the nucleon are determined within the Hessian method at the next-to-next-to leading order (NNLO) in perturbative QCD, based on the most recent measurements from the Large Hadron Collider (LHC) and a variety of world collider data. Because of difficulties in fitting both the ATLAS 7 and 8 TeV $W$ and $Z$ vector boson production cross section data, we present two families of PDFs, named CT18 and CT18$Z$ PDFs, respectively, without and with the ATLAS 7 TeV $W$ and $Z$ measurements. We study the impact of the CT18 family of PDFs on the theoretical predictions of standard candle cross sections at the LHC.}

\dedicated{MSUHEP-19-017, SMU-HEP-19-12}

\FullConference{XXVII International Workshop on Deep-Inelastic Scattering and Related Subjects - DIS2019\\
		8-12 April, 2019\\
		Torino, Italy}

\begin{document}

The CT18 parton distribution functions (PDFs) update those of CT14 presented in Ref.~\cite{Dulat:2015mca} with a variety of new LHC data, involving inclusive jet production, $W$, $Z$ and Drell-Yan production, and the production of top quark pairs, from ATLAS, CMS and LHCb, while retaining crucial {\it legacy} data, such as measurements from the Tevatron and the HERA Run I and Run II combined data. Measurements of processes in similar kinematic regions, by ATLAS and by CMS, allow crucial cross-checks of the data. Measurements by LHCb often allow extrapolations into new kinematic regions not covered by the other experiments. 

The goal of the CT18 analysis is to include as wide a kinematic range for each measurement as allowed by reasonable agreement between data and theory. For the ATLAS 7 TeV jet data~\cite{Aad:2014vwa}, for example, all rapidity intervals can not be simultaneously used without the use of systematic error decorrelations provided by the ATLAS experiment. 
Even with the ATLAS-recommended decorrelations, the resultant $\chi^2$ is not optimal, resulting in less effective PDF constraints.
Inclusive cross section measurements for jet production have been carried out for two different jet radii by both ATLAS and CMS. For both experiments, we have chosen the data with the larger $R$-value, as the next-to-next-to-leading order (NNLO) prediction should have a higher accuracy. We evaluate the jet cross section predictions using a scale of $p_T^{jet}$, consistent with past usage at the next-to-leading order (NLO). The result is largely consistent with similar evaluations using a scale of $H_T$~\cite{Ridder:2015dxa}.

In an ideal world, all such data sets would perfectly be compatible with each other, but differences are observed that do result in some tension between data sets and pulls in opposite directions. One of the crucial aspects of carrying out a global PDF analysis is dealing with data sets that add some tension to the fits, while preserving the ability of the combined data set to improve on the existing constraints on the PDFs. In some cases, a data set may be in such tension as to require its removal from the global analysis, or its inclusion only in a separate iteration of the new PDF set. Later, we will describe how the high precision ATLAS $W$ and $Z$ rapidity distributions \cite{Aaboud:2016btc} require the latter treatment.

Theoretical predictions for comparison to the data used in the global fit have been carried out at NNLO, either indirectly through the use of fast interpolation tables such as NLO \textsc{ApplGrid}~\cite{Carli:2010rw} and NNLO/NLO $K$-factors, or directly (for top-quark related observables) through the use of \textsc{fastNNLO} grids~\cite{Czakon:2017dip,Wobisch:2011ij}. New flexible PDF parametrizations have been tested for CT18, to minimize any parametrization bias. In some kinematic regions, there are few constraints from the data on certain PDFs. Lagrange Multiplier constraints are then applied to limit those PDFs to physically reasonable values. 
A 0.5\% uncorrelated error is included to account for numerical uncertainties in the Monte Carlo integration of NNLO cross sections of (i) ATLAS 7 TeV ~\cite{Aad:2014vwa} and CMS 7~\cite{Chatrchyan:2014gia} and 8 TeV~\cite{Khachatryan:2016mlc} jet productions; and (ii) ATLAS 8 TeV high-$p_T$ $Z$ production~\cite{Aad:2015auj}.

CT18 analysis includes new LHC experiments on $W$, $Z$, Drell-Yan, high-$p_T$ $Z$, jet, and $t \bar t$ pair productions, up to 30 candidate LHC data sets. 
The alternative CT18Z fit contains the following variations from the CT18 fit: (i) add in the ATLAS 7 TeV 4.6 fb$^{-1}$, $W$ and $Z$ rapidity distribution measurement~\cite{Aaboud:2016btc} which is not included in the CT18 fit, (ii) remove the CDHSW data, (iii) take charm pole mass to be 1.4 GeV, instead of the nominal value of 1.3 GeV, (iv) use a saturation scale, instead of the nominal scale of $Q$, for all the deep-inelastic scattering (DIS) processes in the fit.  
The final CT18(Z) data ensemble contains a total of 3681(3493) number of data points and $\chi^2/N_{pt}=1.17 (1.19)$ at the NNLO. 

\begin{figure}[tb]
	\center
	\includegraphics[width=0.43\textwidth]{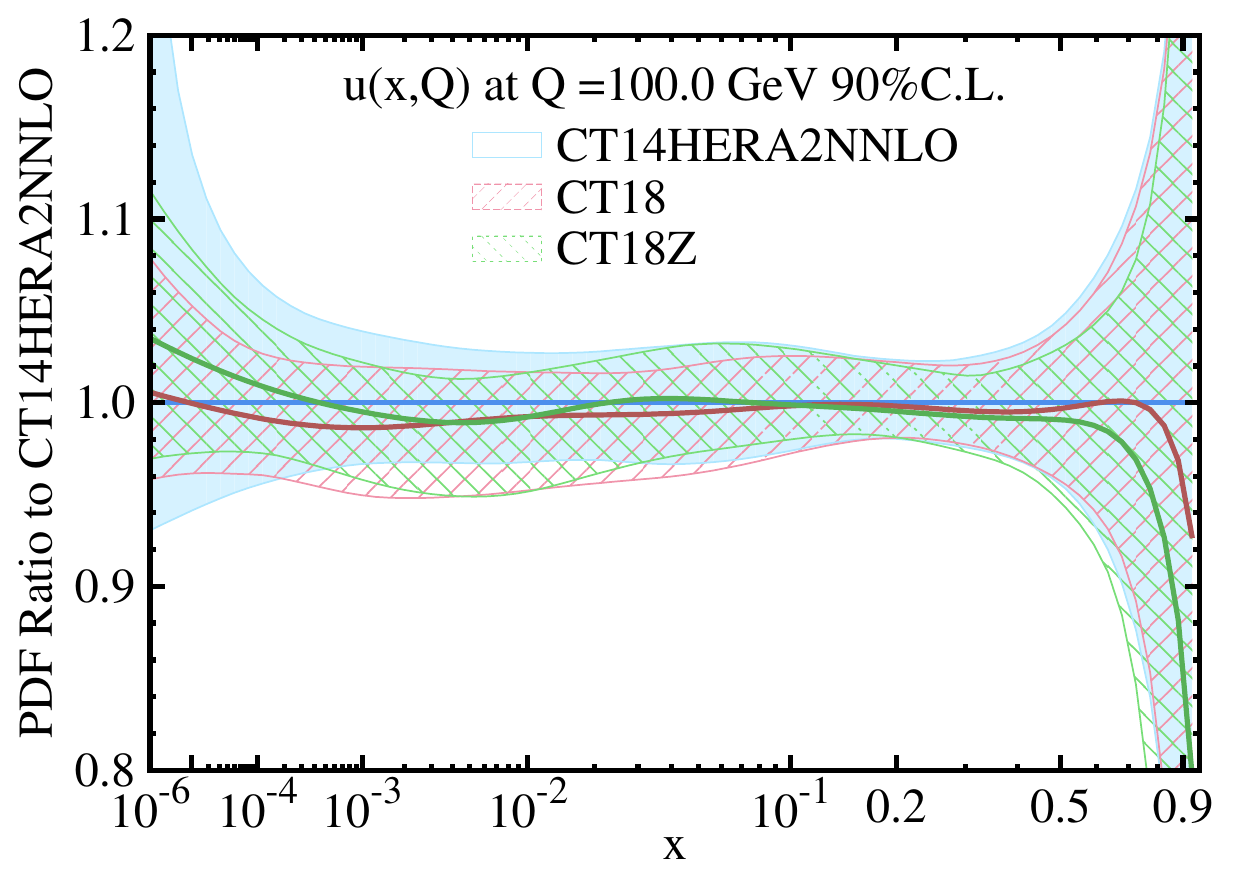}
	\includegraphics[width=0.43\textwidth]{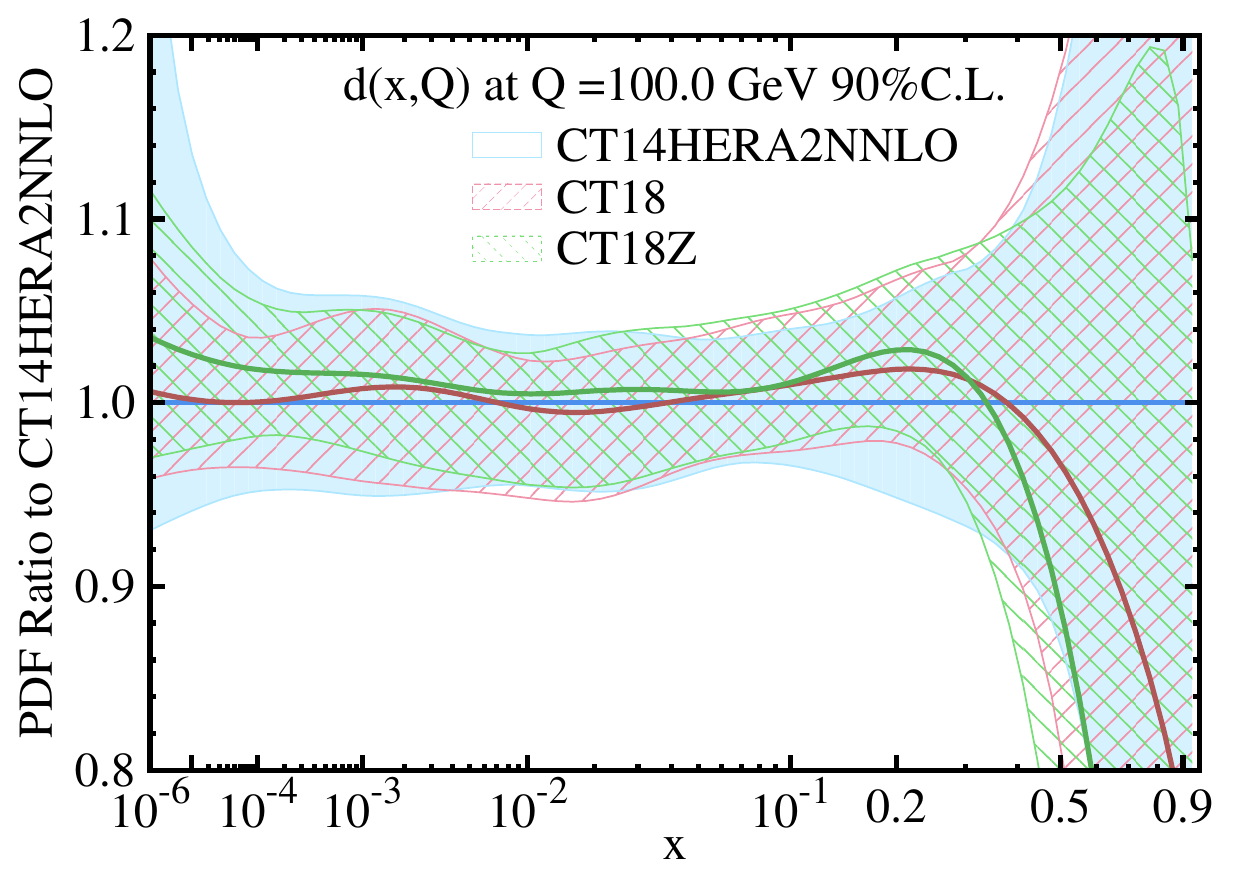}
	\includegraphics[width=0.43\textwidth]{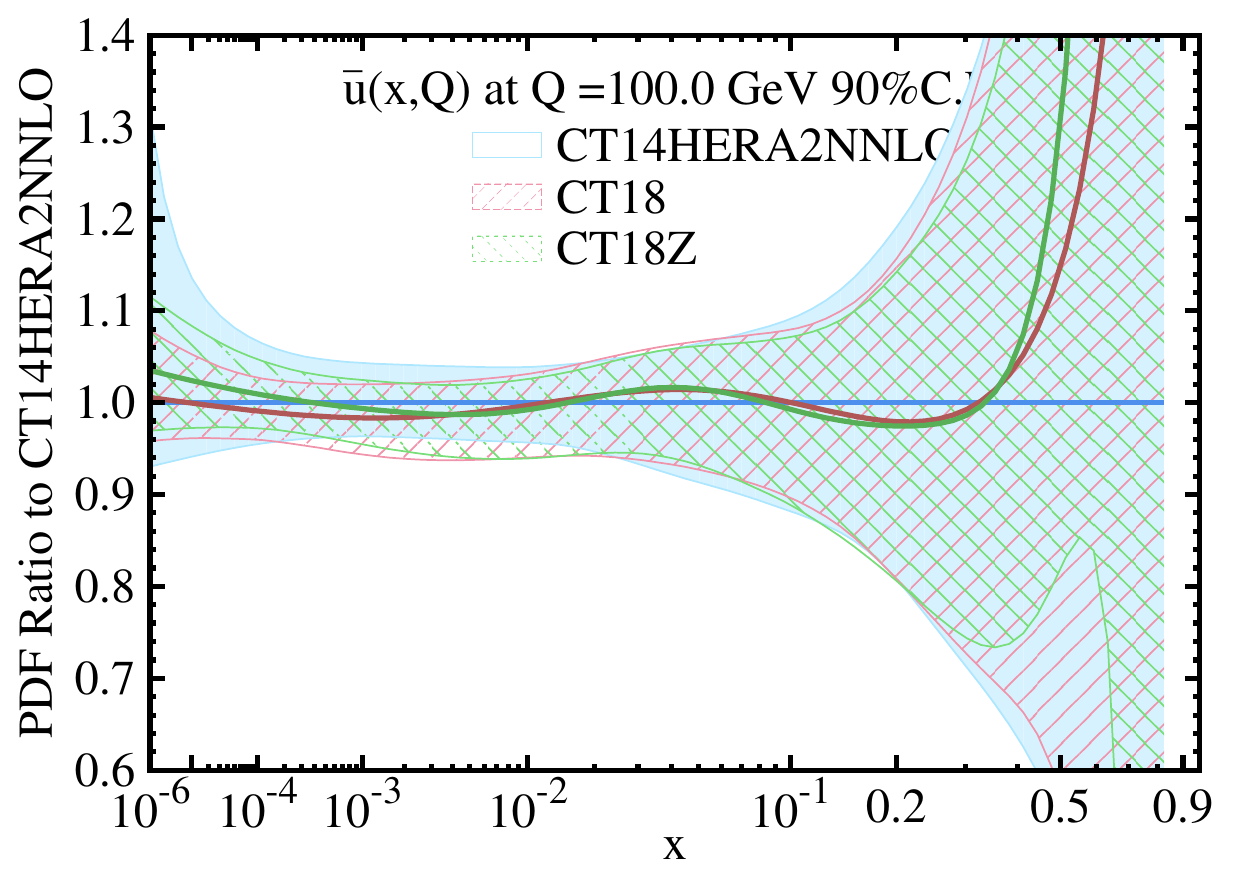}
	\includegraphics[width=0.43\textwidth]{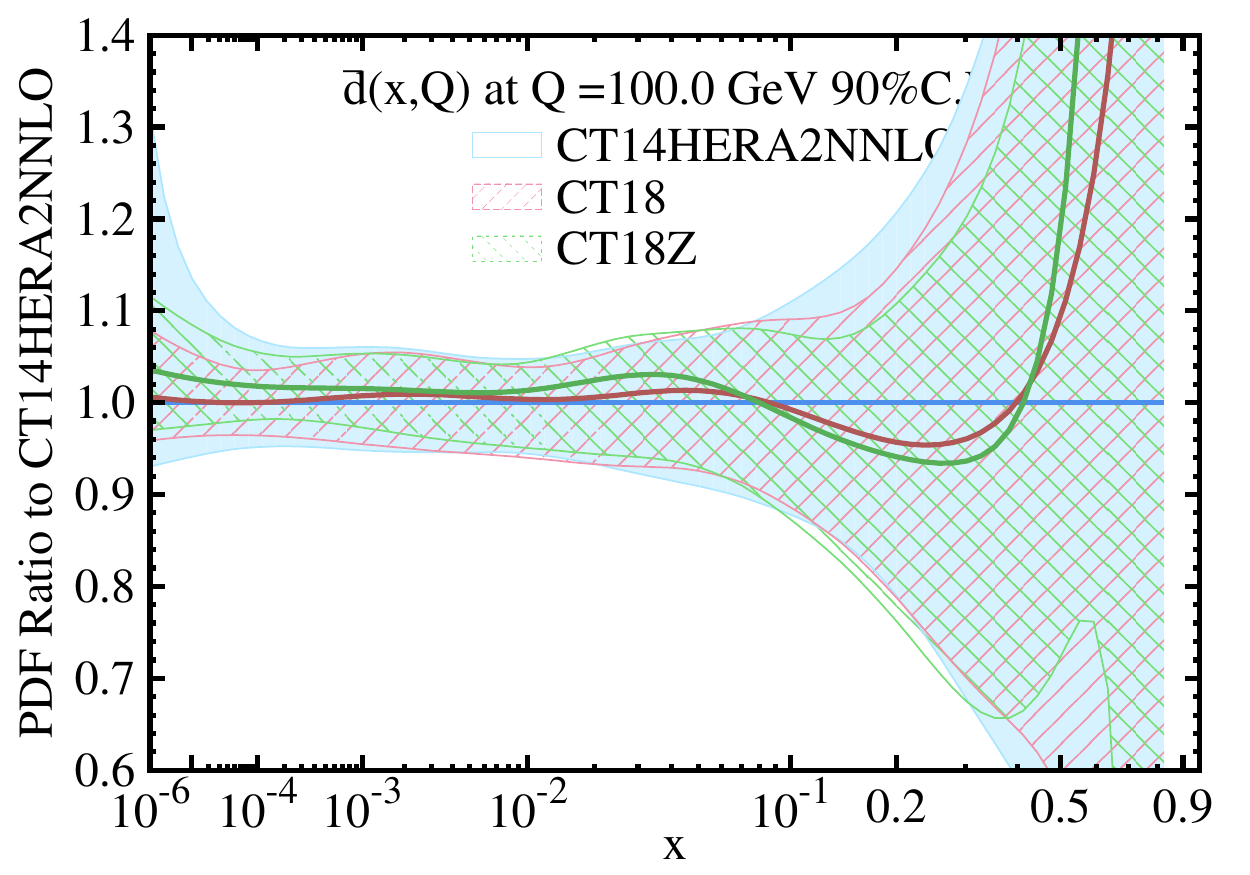}
	\includegraphics[width=0.43\textwidth]{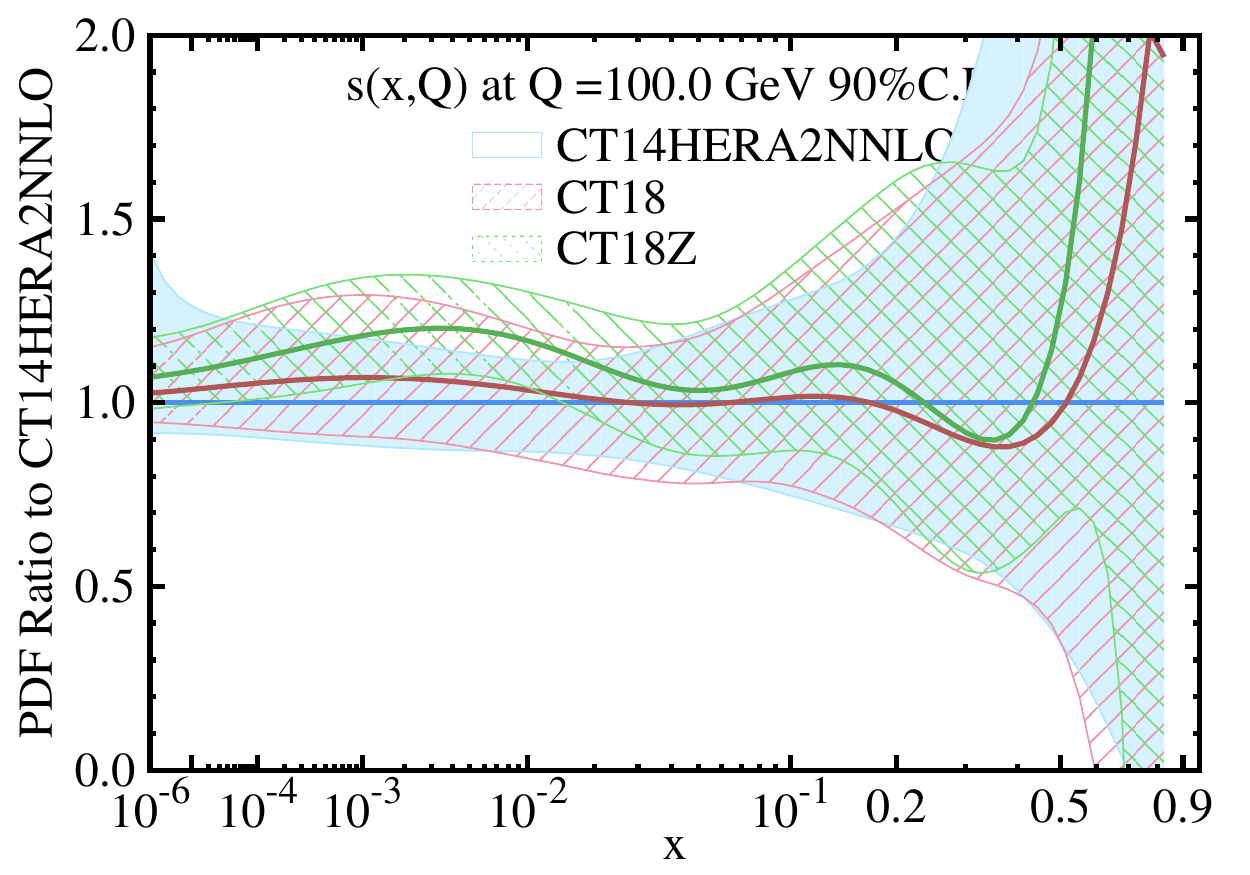}
	\includegraphics[width=0.43\textwidth]{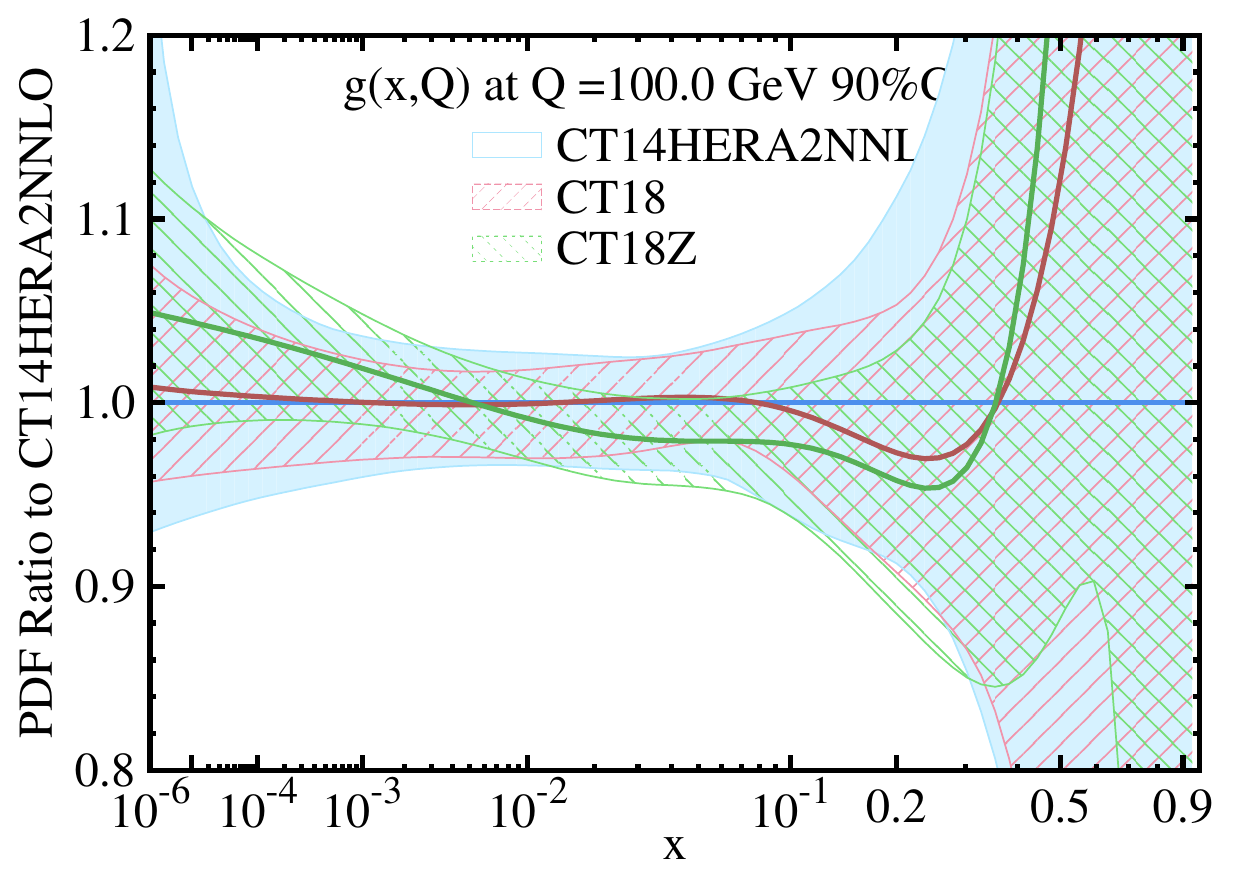}
	\caption{A comparison of 90\% C.L. PDF uncertainties from CT18  (red curve), CT18Z (green curve) and CT14HERA2 (blue curve) NNLO error ensembles at $Q=100$ GeV. The error bands are
		normalized to the respective central CT14HERA2 NNLO PDFs.
		\label{fig:PDFbands1}}
\end{figure}

The relative changes between the CT14HERA2
NNLO and CT18 NNLO ensembles are best visualized by comparing
their PDF uncertainties. Fig.~\ref{fig:PDFbands1}
compares the PDF error bands at 90\% confidence level (CL)
for the key flavors, with each band normalized to the respective
best-fit CT14HERA2 NNLO PDF. The blue and red error bands are
obtained for CT14HERA2 NNLO PDFs and CT18 at $Q=100$ GeV, respectively.

\begin{figure}[tbp]
	\center
	\includegraphics[width=0.43\textwidth]{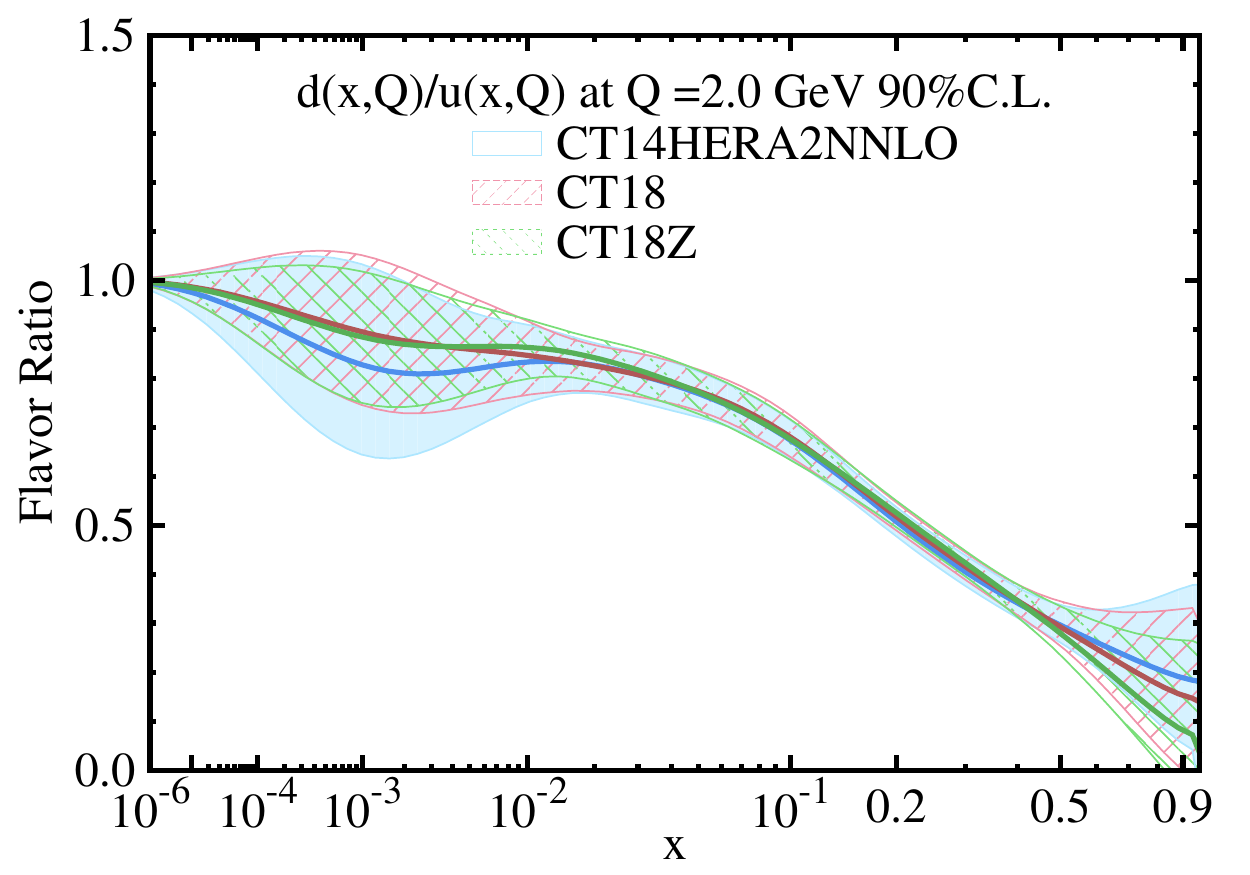}
	\includegraphics[width=0.43\textwidth]{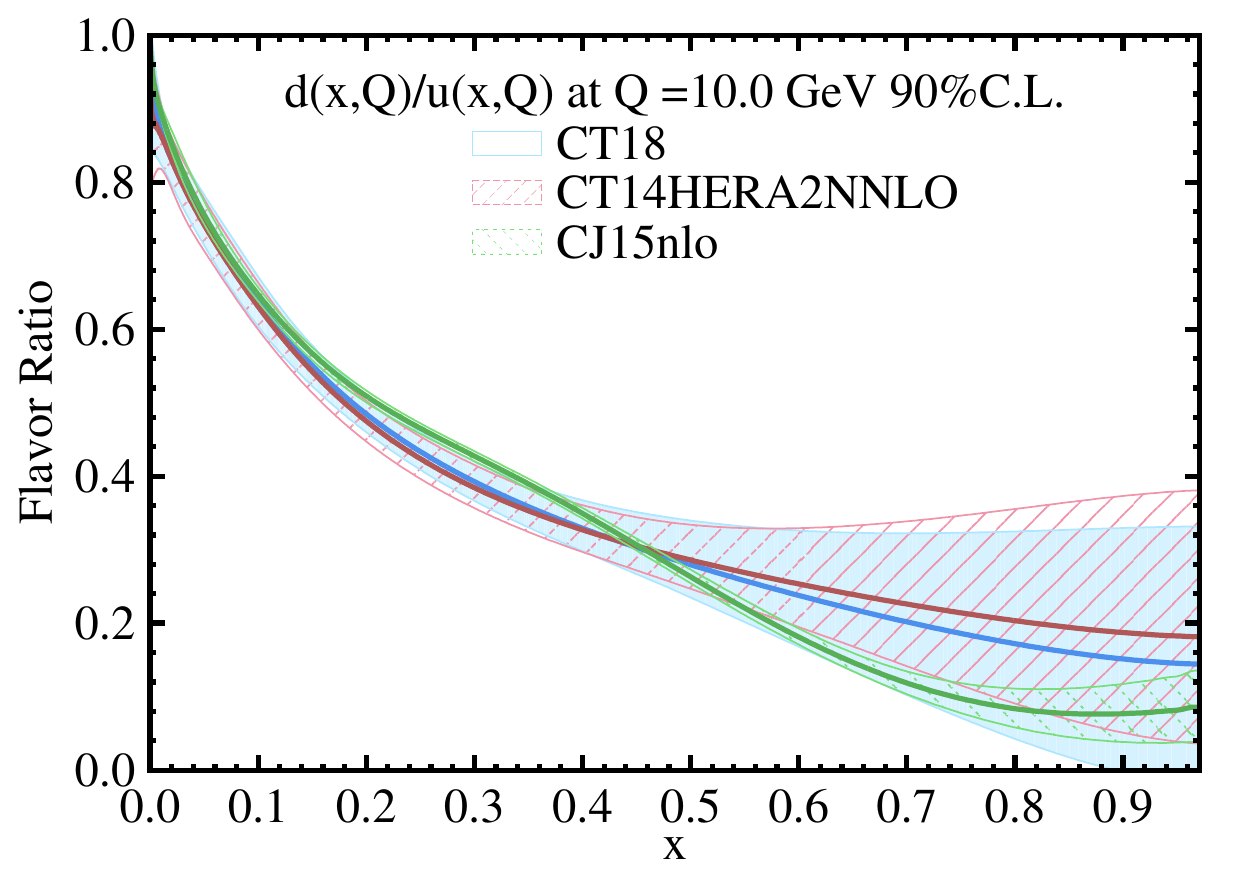}
	\caption{A comparison of 90\% C.L. uncertainties on the ratio
		$d(x,Q)/u(x,Q)$ for CT18  (red curve), CT18Z (green curve) and CT14HERA2 (blue curve) NNLO error ensembles at $Q=2$ or $10$ GeV, respectively.
		\label{fig:DOUband}}
\end{figure}

\begin{figure}[tb]
	\center
	\includegraphics[width=0.43\textwidth]{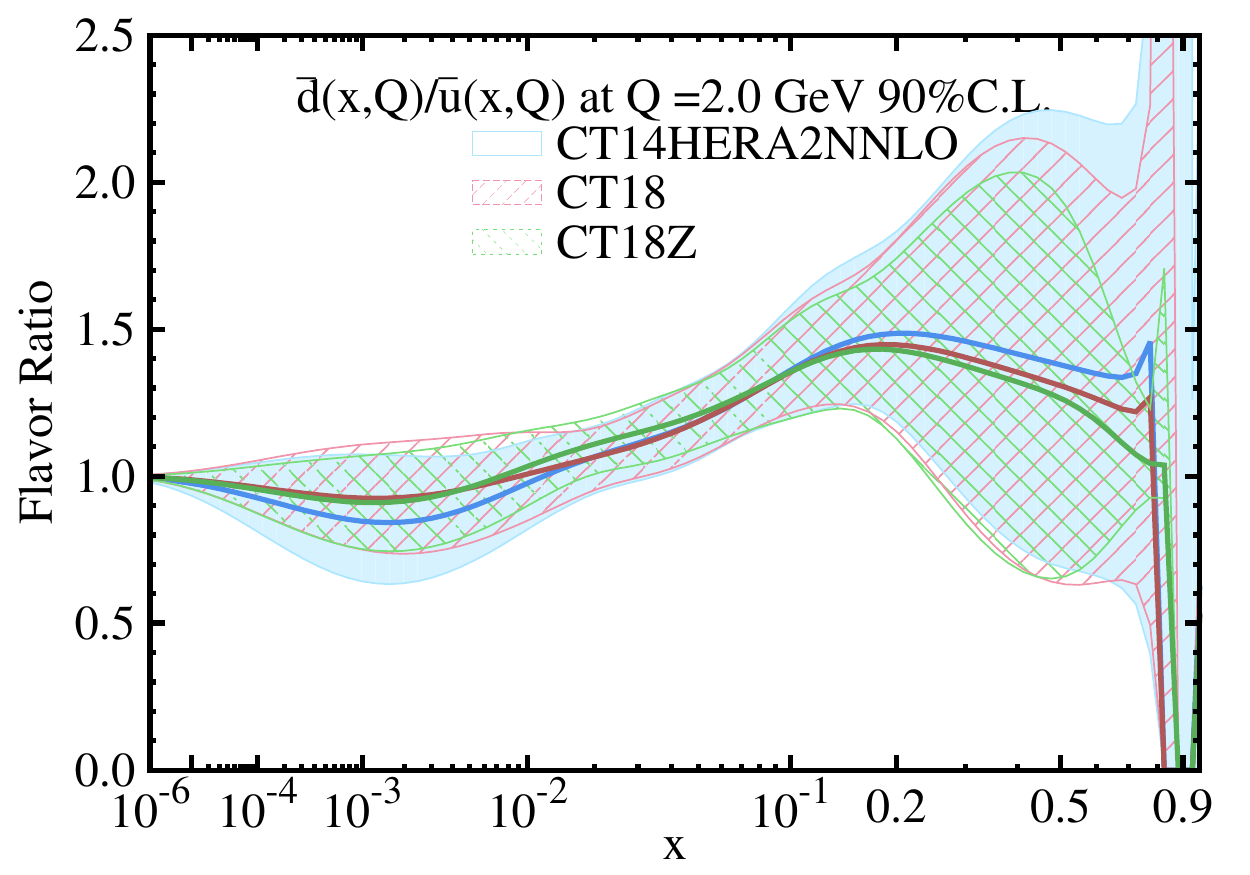}
	\includegraphics[width=0.43\textwidth]{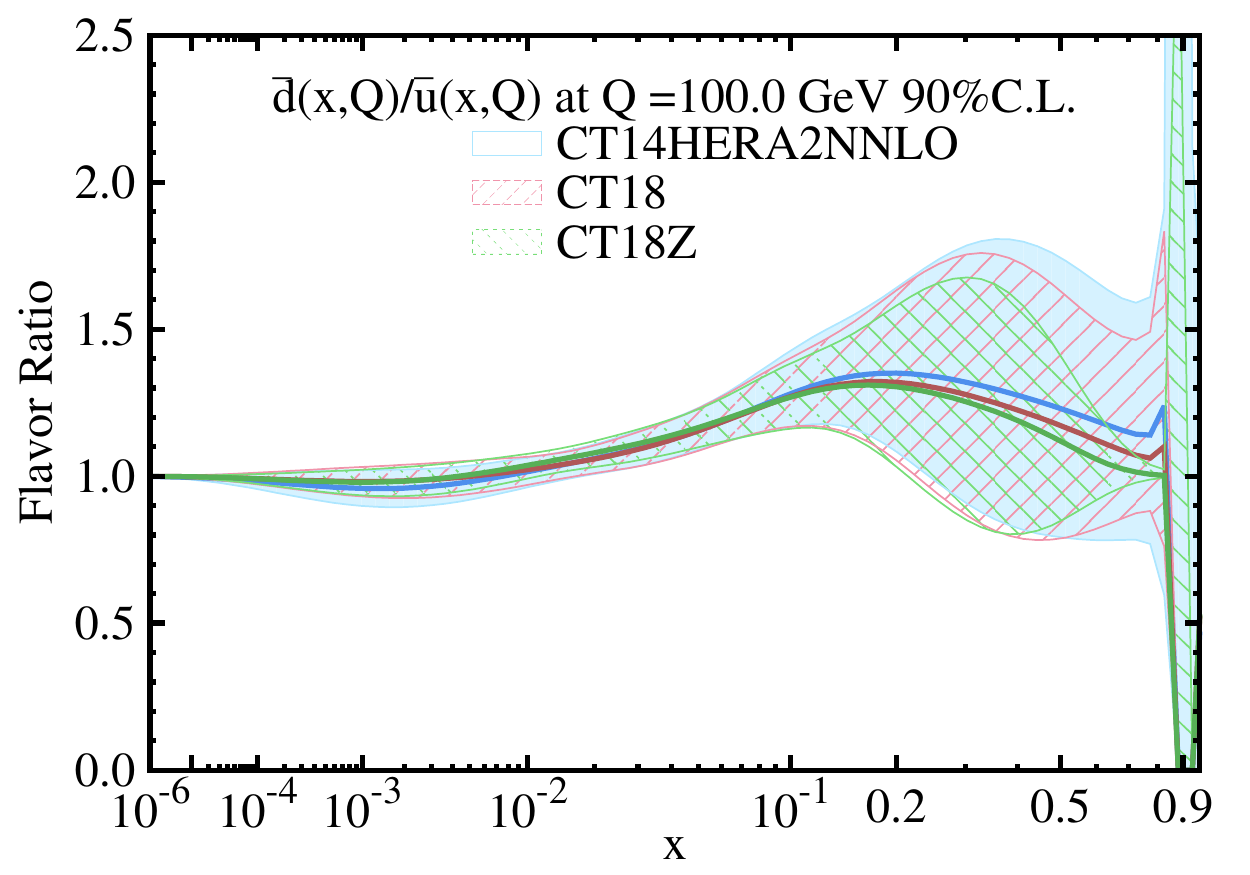}
	\includegraphics[width=0.43\textwidth]{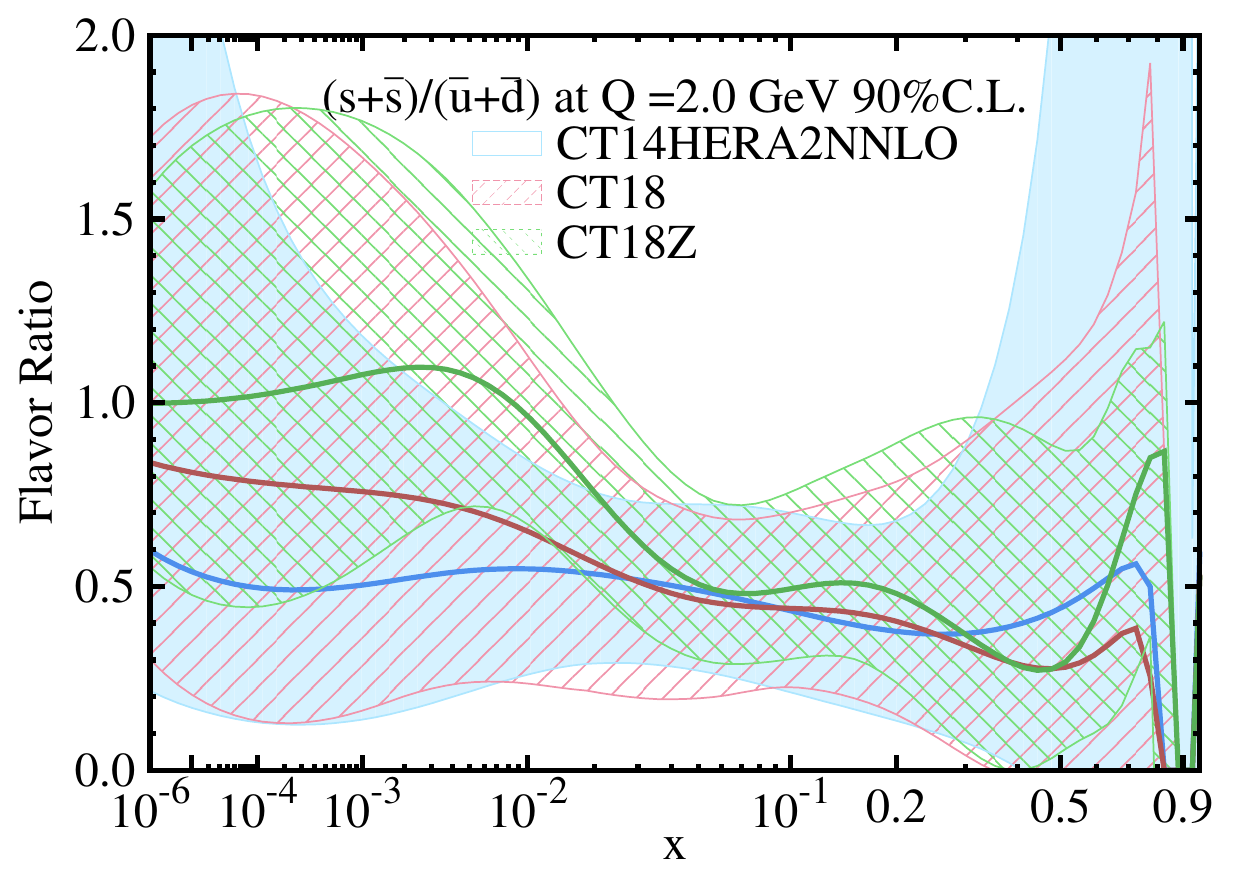}
	\includegraphics[width=0.43\textwidth]{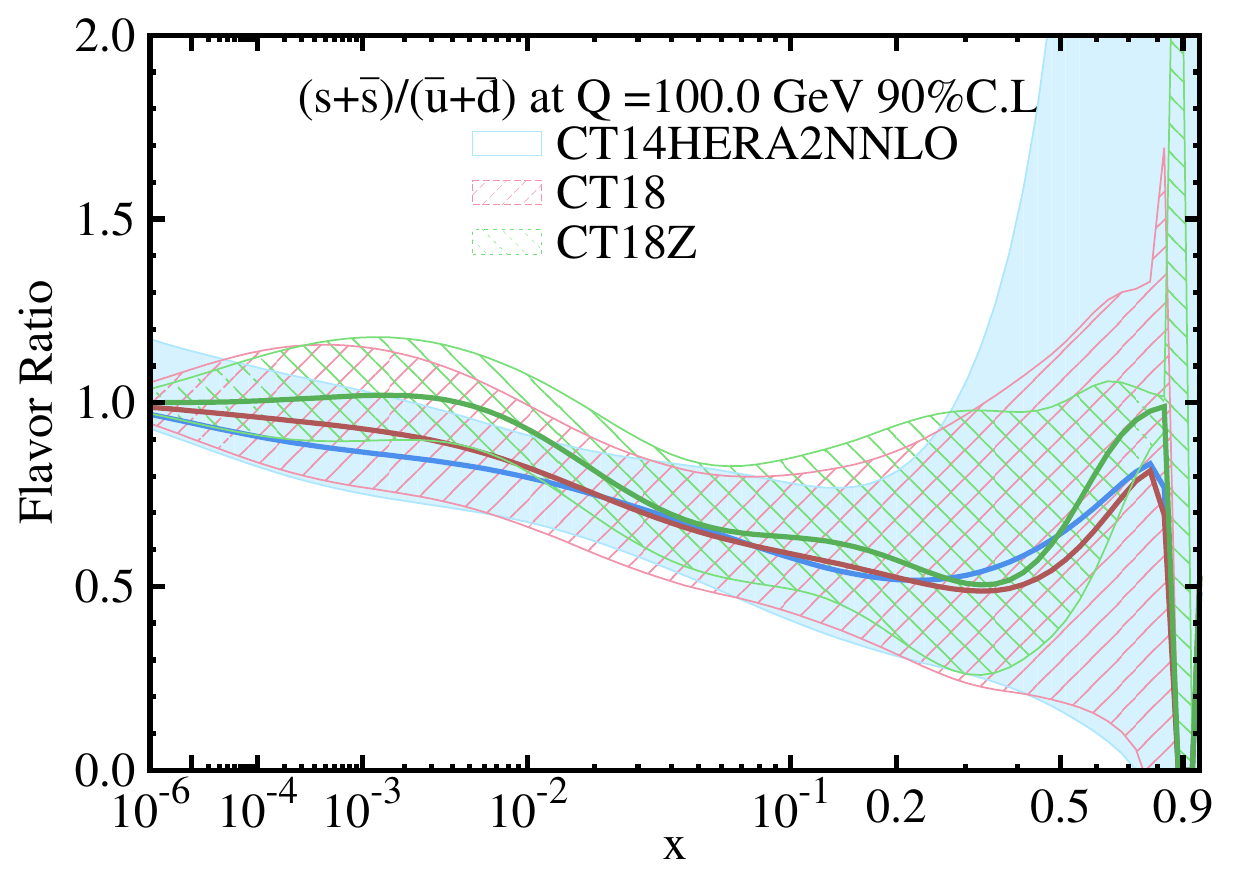}
	\caption{A comparison of 90\% C.L. uncertainties on the ratios
		$\bar d(x,Q)/\bar u(x,Q)$ and $\left(s(x,Q)+\bar
		s(x,Q)\right)/\left(\bar u(x,Q) +\bar d(x,Q)\right)$, for CT18  (red curve), CT18Z (green curve) and CT14HERA2 (blue curve) NNLO error ensembles at $Q=2$ or $100$ GeV, respectively.
		\label{fig:DBandSBbands}}
\end{figure}

Fig.~\ref{fig:DOUband} shows the different behaviors we find for the $d/u$ PDF ratio.
The changes in $d/u$ in CT18, as compared to CT14HERA2, can be summarized as a reduction of the central ratio at $x > 0.5$ and an
decreased uncertainty at $x < 10^{-3}$.
The collider charge asymmetry data constrains $d/u$ at $x$ up to about 0.5. 
At even higher $x$, which is not directly constrained by the experiments we fit, the behavior of the CT18 PDFs reflects the parametrization form,
which now allows $d/u$ to
approach any constant value as $x\rightarrow 1$.
As noted earlier, the parametrization form of $u$, $d$, $\bar u$ and $\bar d$ quarks in CT18 are the same as those in CT14HERA2. 

Turning now to the ratios of sea quark PDFs in
Fig.~\ref{fig:DBandSBbands}, we observe that the uncertainty on $\bar
d(x,Q)/\bar u(x,Q)$ in the left inset has decreased at small $x$
in CT18. At $x > 0.1$, the CT18 non-perturbative parametrization forms for $\bar u$ and $\bar d$ ensure that the ratio
$\bar d(x,Q_0)/\bar u(x,Q_0)$, with $Q_0=1.3$ GeV, can approach a constant value that
comes out to be close to 1 in the central fit. The uncertainty on
$\bar d/\bar u$ has also decreased across most of the $x > 2 \times 10^{-3}$ range, especially around $x \sim 0.1$.

The overall increase in the strangeness PDF at $ x < 0.03$ and decrease of $\bar u$ and 
$\bar d$ PDFs at $ x  < 10 ^{-3}$ lead to a larger ratio of the strange-to-nonstrange sea quark PDFs,
$\left(s+\bar s\right)/\left(\bar u +\bar d \right)$, presented in
Fig.~\ref{fig:DBandSBbands}. 
At $ x  < 10 ^{-3}$, this
ratio is determined entirely by parametrization form and was found
in CT10 to be consistent with the exact $SU(3)$ symmetry of PDF
flavors, 
$\left(s+\bar s\right)/\left(\bar u +\bar d\right) \rightarrow 1$
at $x\rightarrow 0$, albeit with a large uncertainty. 
The $SU(3)$-symmetric asymptotic solution at $x\rightarrow 0$
was not enforced in CT14, nor CT14HERA2, so that this ratio at $Q=2$ GeV 
is about 0.6 at $x=10^{-6}$. In CT18, we have taken a different $s$-PDF non-perturbative parametrization form and assumed the exact $SU(3)$ symmetry of PDF
flavors so that this ratio asymptotically approaches to 1 as $x \to 0$.

\begin{figure}[h]
	\includegraphics[width=0.48\textwidth]{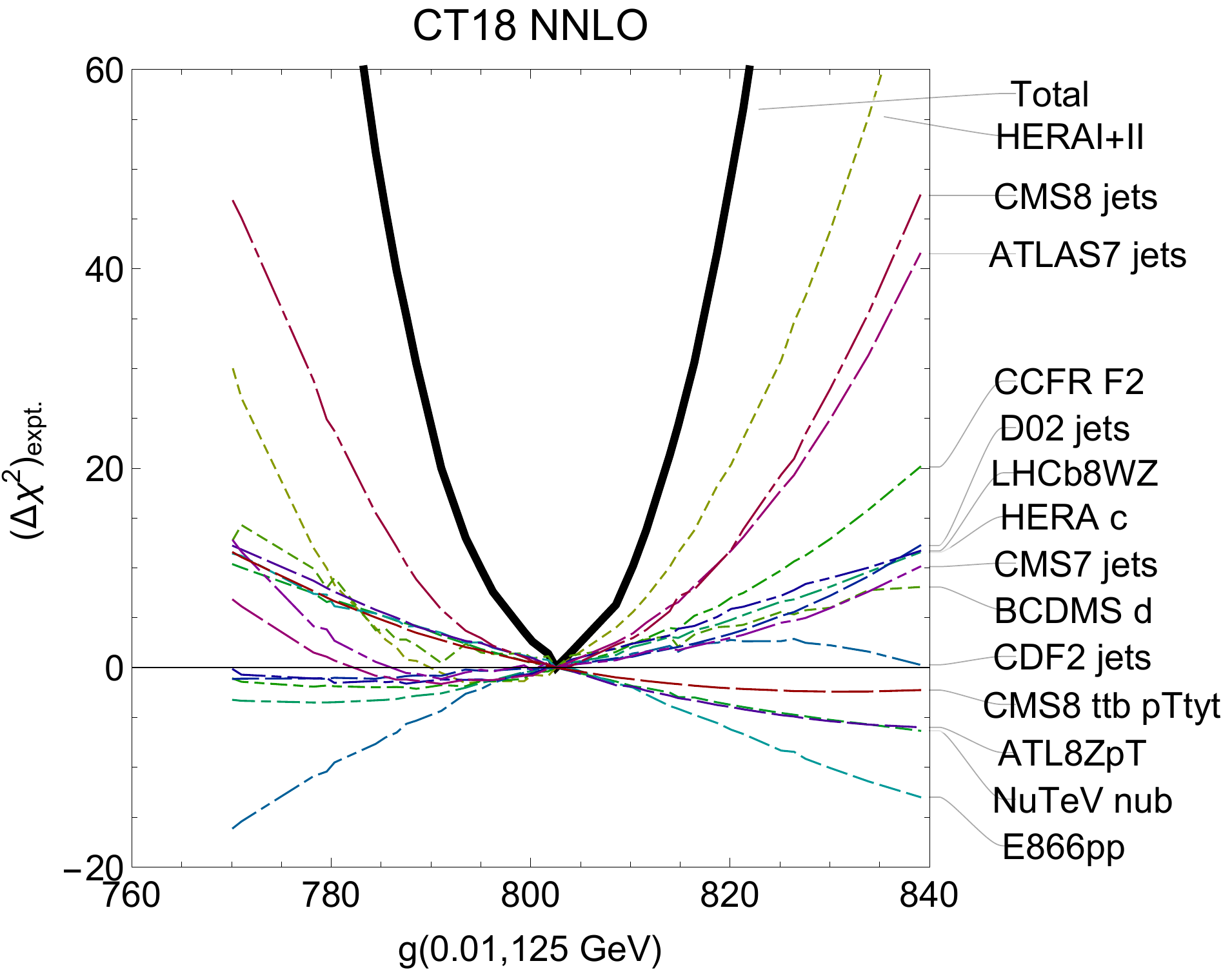}\quad
	\includegraphics[width=0.48\textwidth]{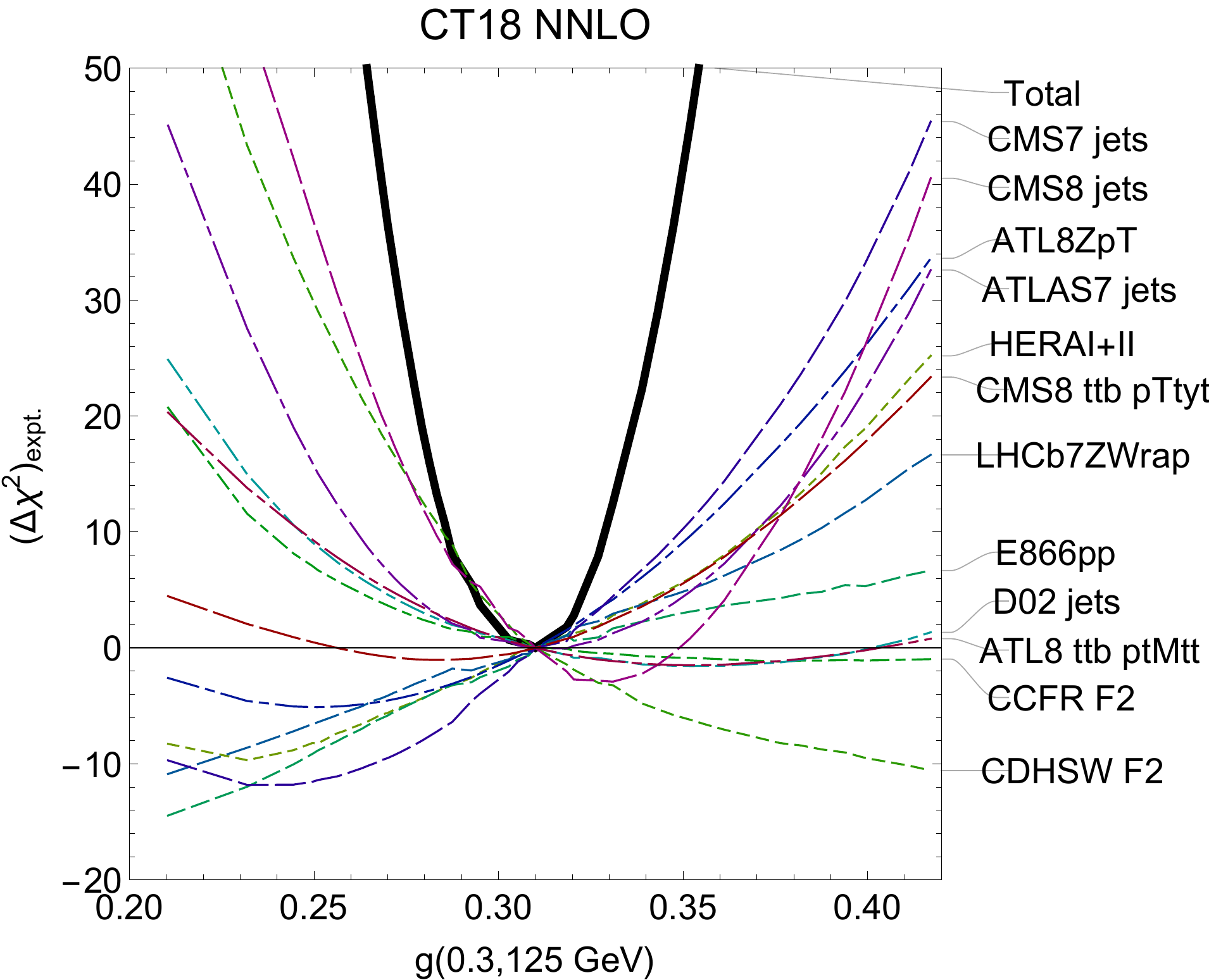}
	\caption{The Lagrange Multiplier scan of gluon PDF at $Q=125$ GeV and $x=0.01$ and $0.3$, respectively, for the CT18 NNLO fits.
		\label{fig:lm_g18}}
\end{figure}

\begin{figure}[h]
	\begin{center}
		\includegraphics[width=0.48\textwidth]{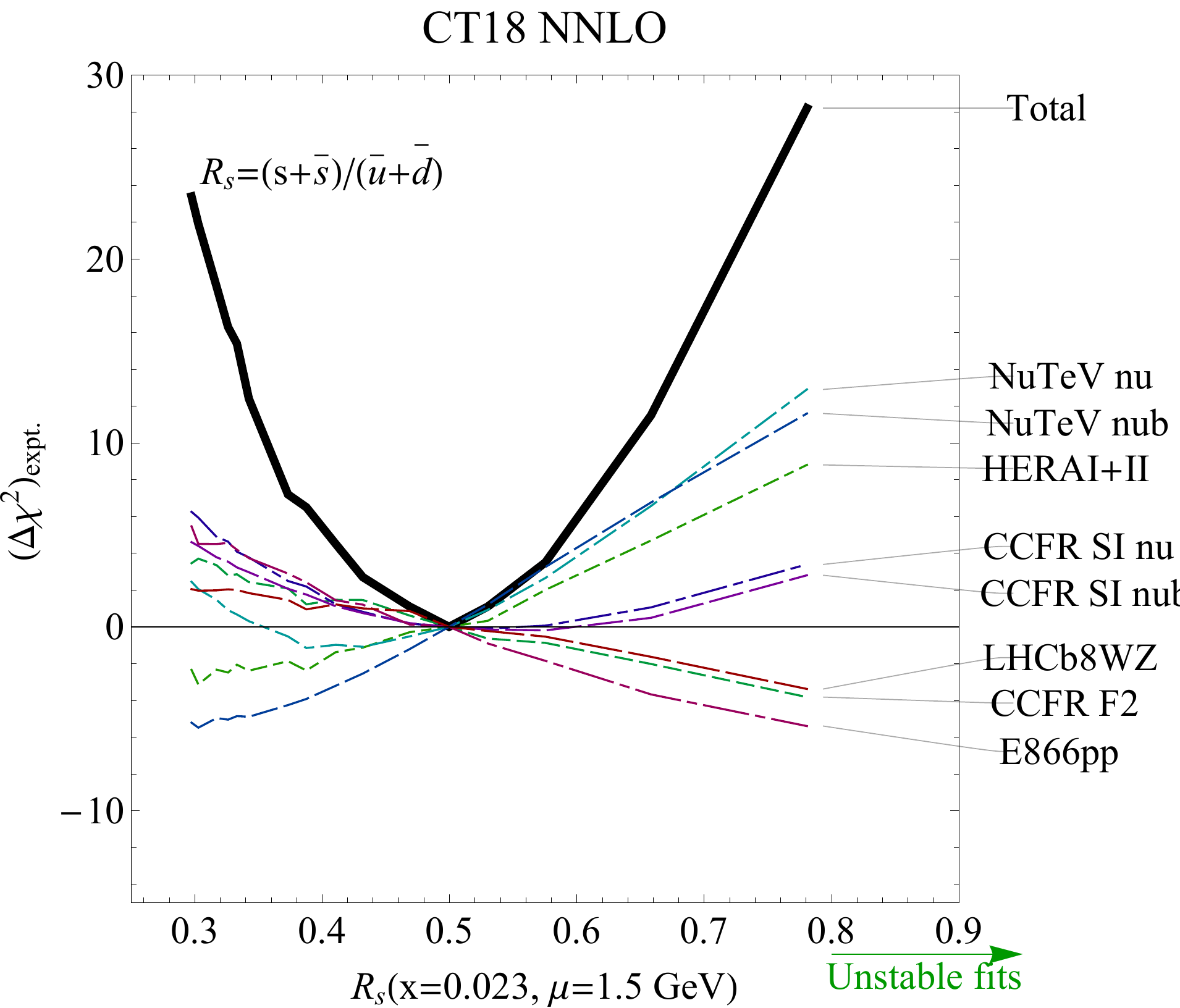}\quad
		\includegraphics[width=0.48\textwidth]{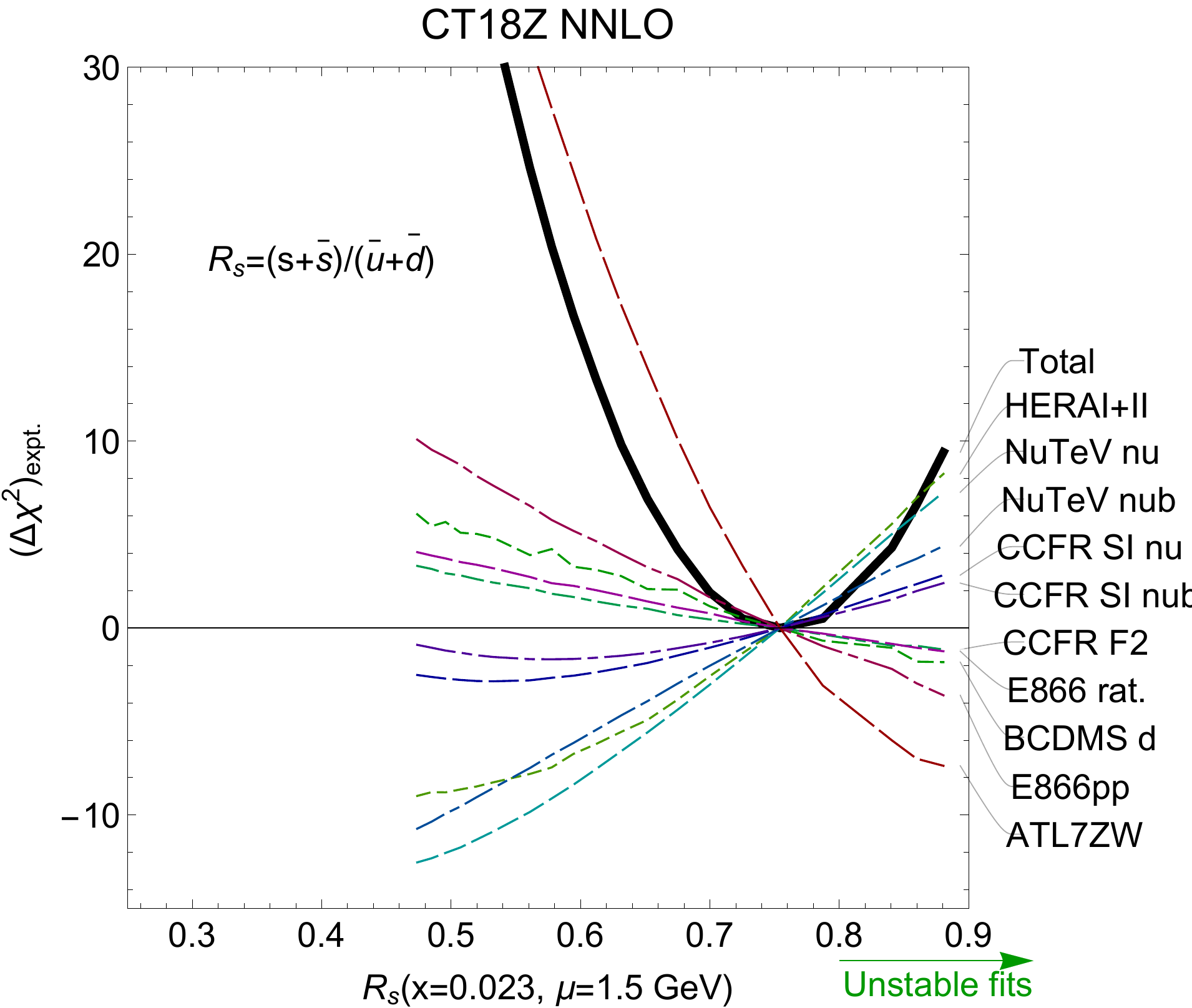}\quad
		\caption{The Lagrange Multiplier scan of $R_s$ at $Q=1.5$ GeV and $x=0.023$ for CT18, and CT18Z fits.
			\label{fig:lm_rs}}
	\end{center}
\end{figure}

\begin{figure}[h]
	\begin{center}
		\includegraphics[width=0.45\textwidth]{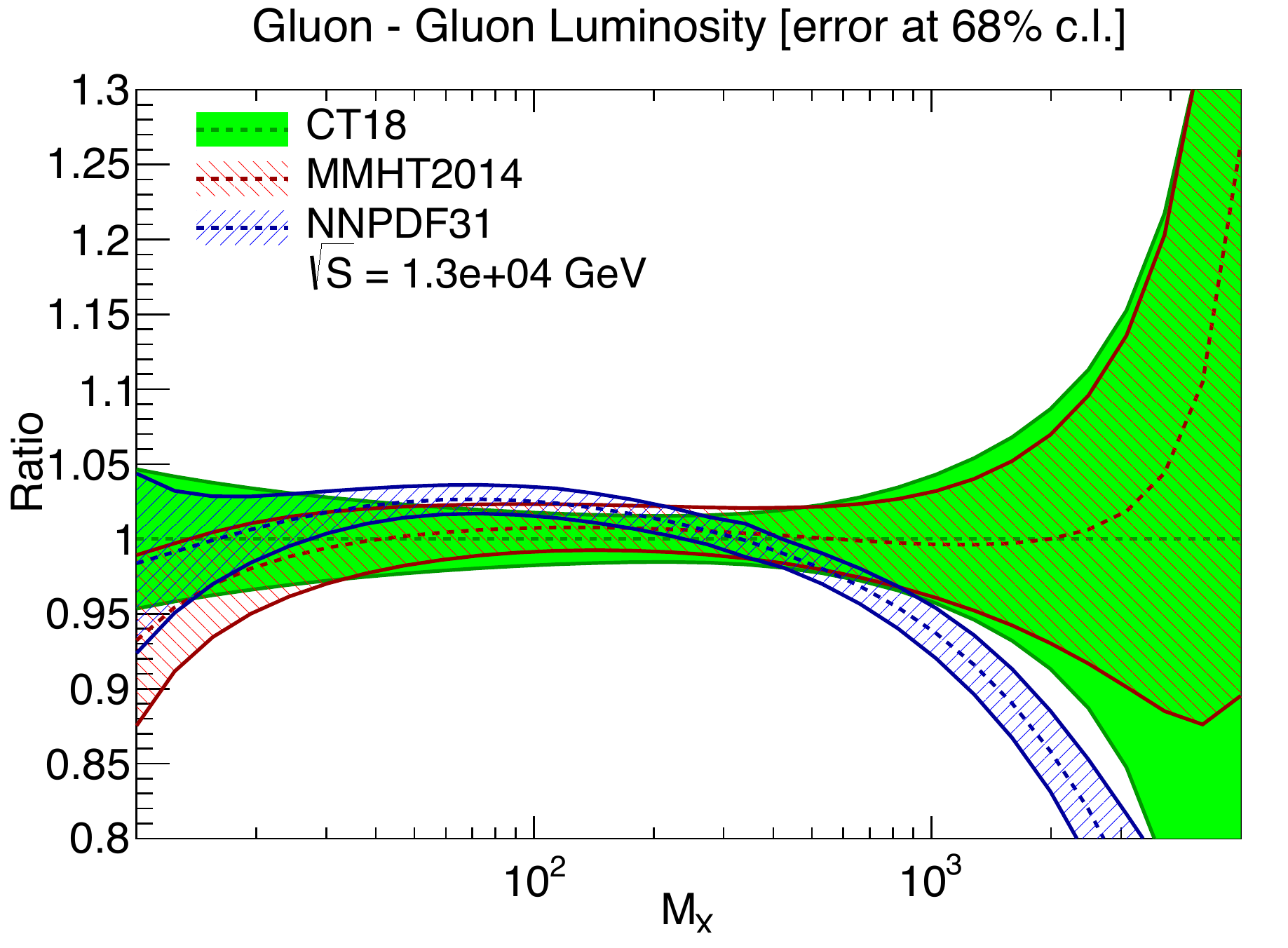}
		\includegraphics[width=0.45\textwidth]{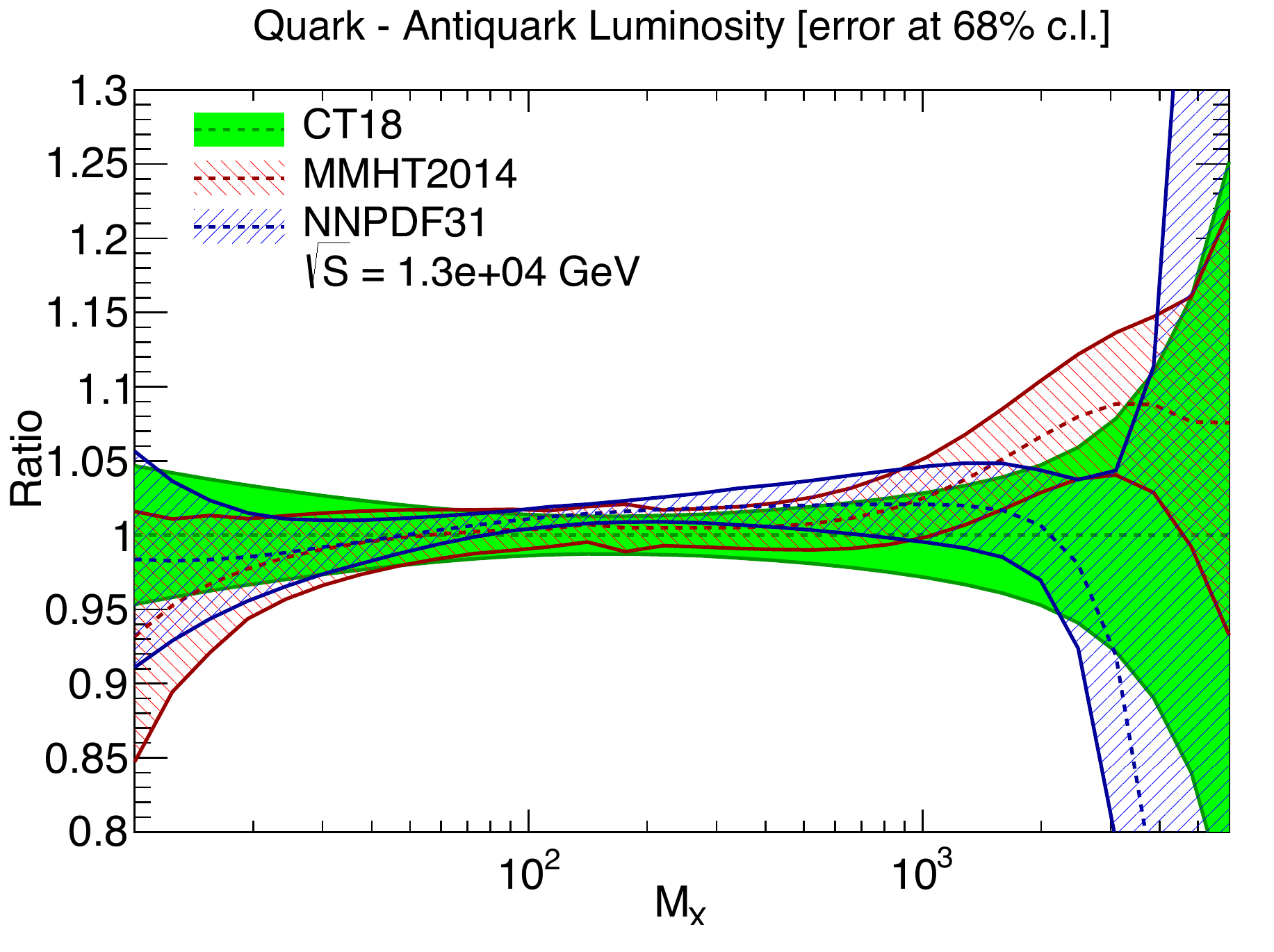}		
	\end{center}
	\vspace{-2ex}
	\caption{\label{fig:lumib}
		Comparison of various PDF luminosities at the 13 TeV LHC. 
	}
\end{figure}

One technique that we use to study the parton PDFs is to compute Lagrange Multiplier scans with respect to some feature of $f(x,Q)$. Two examples are shown here. 
First example is to study the constraints on gluon-PDF at $Q=125$ GeV and $x=0.01$ and $0.3$, from various experimental data, cf. Fig.~\ref{fig:lm_g18}. 
The second example is for the constraints on 
the $R_s \equiv (s+ \bar s)/({\bar u}+ {\bar d})$ ratio at $Q=1.5$ GeV, $x=0.023$ and $x=0.1$, cf. Fig.~\ref{fig:lm_rs}. 

Finally, we compare various PDF luminosities at the 13 TeV LHC, as shown in Fig.~\ref{fig:lumib}.

\begin{acknowledgments}
The work of J.~Gao was sponsored by the National Natural Science Foundation of China under the Grant No. 11875189 and No.11835005. 
The work at SMU is supported by the U.S. Department of Energy under Grant No. DE-SC0010129.
The work of M.G. is supported by the National Science Foundation under Grant No. PHY1820818.
The work of C.-P. Yuan was supported by the U.S. National Science Foundation under Grant No. PHY-1719914, and he is also grateful for the support from the Wu-Ki Tung endowed chair in particle physics.	
	
\end{acknowledgments}


\begin{thebibliography}{99}


\bibitem{Dulat:2015mca} 
S.~Dulat {\it et al.},
Phys.\ Rev.\ D {\bf 93}, no. 3, 033006 (2016)
doi:10.1103/PhysRevD.93.033006
[arXiv:1506.07443 [hep-ph]].


\bibitem{Aad:2014vwa} 
G.~Aad {\it et al.} [ATLAS Collaboration],
JHEP {\bf 1502}, 153 (2015)
Erratum: [JHEP {\bf 1509}, 141 (2015)]
doi:10.1007/JHEP02(2015)153, 10.1007/JHEP09(2015)141
[arXiv:1410.8857 [hep-ex]].


\bibitem{Ridder:2015dxa} 
A.~Gehrmann-De Ridder, T.~Gehrmann, E.~W.~N.~Glover, A.~Huss and T.~A.~Morgan,
Phys.\ Rev.\ Lett.\  {\bf 117}, no. 2, 022001 (2016)
doi:10.1103/PhysRevLett.117.022001
[arXiv:1507.02850 [hep-ph]].


\bibitem{Aaboud:2016btc} 
M.~Aaboud {\it et al.} [ATLAS Collaboration],
Eur.\ Phys.\ J.\ C {\bf 77}, no. 6, 367 (2017)
doi:10.1140/epjc/s10052-017-4911-9
[arXiv:1612.03016 [hep-ex]].


\bibitem{Carli:2010rw} 
T.~Carli, D.~Clements, A.~Cooper-Sarkar, C.~Gwenlan, G.~P.~Salam, F.~Siegert, P.~Starovoitov and M.~Sutton,
Eur.\ Phys.\ J.\ C {\bf 66}, 503 (2010)
doi:10.1140/epjc/s10052-010-1255-0
[arXiv:0911.2985 [hep-ph]].


\bibitem{Czakon:2017dip} 
M.~Czakon, D.~Heymes and A.~Mitov,
arXiv:1704.08551 [hep-ph].


\bibitem{Wobisch:2011ij} 
M.~Wobisch {\it et al.} [fastNLO Collaboration],
arXiv:1109.1310 [hep-ph].


\bibitem{Chatrchyan:2014gia} 
S.~Chatrchyan {\it et al.} [CMS Collaboration],
Phys.\ Rev.\ D {\bf 90}, no. 7, 072006 (2014)
doi:10.1103/PhysRevD.90.072006
[arXiv:1406.0324 [hep-ex]].


\bibitem{Khachatryan:2016mlc} 
V.~Khachatryan {\it et al.} [CMS Collaboration],
JHEP {\bf 1703}, 156 (2017)
doi:10.1007/JHEP03(2017)156
[arXiv:1609.05331 [hep-ex]].


\bibitem{Aad:2015auj} 
G.~Aad {\it et al.} [ATLAS Collaboration],
Eur.\ Phys.\ J.\ C {\bf 76}, no. 5, 291 (2016)
doi:10.1140/epjc/s10052-016-4070-4
[arXiv:1512.02192 [hep-ex]].


\end{thebibliography}
\end{document}